\documentclass[12pt]{article}
\usepackage{epsfig}
\usepackage{german,a4,}
\usepackage{amsmath}
\usepackage{amssymb}
\begin{document}
\begin{center} {\bf  \Large Some remarks on the genesis of
    scalar-tensor theories.}\normalsize \end{center} 
\vspace{1cm}
\begin{center} Hubert Goenner$^{a}$\\Institute for Theoretical Physics\\ Friedrich-Hund-Platz 1\\University of G\"ottingen\\D 37077 G\"ottingen \end{center}

\vspace{9cm} 
\noindent $a)$\\ e-mail: goenner@theorie.physik.uni-goettingen.de
\newpage

\noindent {\bf Abstract}:\\

\noindent Between 1941 and 1962, scalar-tensor theories of gravitation
were suggested four times by different scientists in four different countries. 
The earliest originator, the Swiss mathematician W. Scherrer, was virtually 
unknown until now whereas the chronologically latest pair gave their names 
to a multitude of publications on Brans-Dicke theory. P. Jordan, one
of the pioneers of  quantum mechanics and quantum field theory, and
Y. Thiry, a student of the mathematician Lichnerowicz, known by his
book on celestial mechanics, complete the quartet. Diverse motivations
for and conceptual interpretations of their theories will be discussed
as well as relations among them. Also, external factors like language,
citation habits, or closeness to the mainstream are considered. It
will become clear why Brans-Dicke theory, although structurally a
d\'ej\`a-vu, superseded all the other approaches.   
\newpage
\section{Introduction}
Among alternative theories of gravitation, scalar-tensor theories have
received more attention than others. In publications using the
English language, they are usually narrowed to a particular theory
bearing the name {\em Brans-Dicke theory}, or sometimes, {\em
  Brans-Dicke-Jordan theory}. In the German physics literature, a then current 
name was {\em Jordan's theory} or ``extended theory of gravity'', in the
French {\em Jordan-Thiry} theory. In the following conceptual
description and interpretation of the genesis of scalar-tensor
theories, we concentrate on the differing physical and mathematical
motivations of the proponents of the theory, and on some factors
outside of physics influencing the course of affairs (communications,
distance to mainstream research, language skills, citation habits). It is
interesting that all four scientists who suggested a scalar-tensor theory 
started from diverse vantage points and gave the scalar field differing 
interpretations.  A methodical divide between these proposals is the choice 
between a projection from five-dimensional space and restriction to four-dimensional 
space-time from the outset. My special interest is to find out why the approach  
by Brans and Dicke succeeded to become a synonym for scalar-tensor
theories (cf. \cite{Kaiser2007}). If  Nordstr\"om's relativistic scalar theory 
of 1913 is put aside \cite{Nordst1913}, the nascency of scalar-tensor theories 
of gravitation has not been studied in any detail until now \cite{CoqEsp1990}, 
\cite{FuMa2003}. This includes the report given by C. Brans himself 
\cite{Brans2005}.\\   

There is a surprisingly large span of time concerning the publication
of the four proposals: from 1941 to 1961. One should think that at the beginning
of the 60s, i. e., about 20 years after the first proposal (Scherrer), and more
than ten years since the next two initiators (Jordan, Thiry) had made
their work public, the main previous approaches to scalar-tensor
theory would have been registered in the community of
relativists. In Princeton, seemingly only Jordan's theory had
become known, not Thiry's investigations in Paris, although these two groups
had been citing each other. Moreover, chronologically, the main
reference in the Princeton publications was to the second edition of
Jordan's book published in 1955, with further references to papers of
his group of the same and later years. But the essential results (and
even more general ones than were reached by Brans-Dicke theory) had been
published already until 1952, when the first edition of Jordan's book
had come out. A conjecture would be that the interruption of
communications during and right after the second world war has played
a role. The unavailability of the first edition and of some of the less known  
German-language journals in which Jordan and his group in Hamburg published, 
may have hindered the spread of knowledge about his scalar-tensor
theory. Also, the current citations habits in the physics community focusing more on recentness than on completeness of the literature referred to may have
contributed. Moreover, contrary to the belief of some historians of science,
i.e., that communication among scientists functions rapidly independently of
the language used, it seems plausible that missing language skills, in
this case mainly of French, are to be taken into account. 

\section{From unified field theory to alternative gravitation} 

\subsection{Jordan and projective relativity}
Towards the end of world war II, Pascual Jordan (1902-1980), one of the 
pioneers of quantum mechanics, had thought about processes by which stars 
are generated and about the time scale of their subsequent evolution.  
This seems to have played a major role for the building of a theory by
him in which the gravitational constant $\kappa= \frac{8\pi G}{c^2}$
is thought to be varying in (cosmological) time and is replaced by a
scalar function. His starting point was Kaluza's unified field theory
in a 5-dimensional space which formally had been rewritten as
projective relativity in space-time (Veblen, Pauli). Its fifteenth
field variable, a scalar which had been set constant, now was seen to
nicely fit to Dirac's ``large number hypothesis'', i. e., his idea that
the fundamental constants (including the age of the universe) might be 
variable in cosmological time \cite{Dirac1937}, \cite{Dirac1938}. 

In his first publication right after the end of the second world war,\footnote{Due
to the warfare afflicting Germany and Europe at the time, Jordan's first paper 
on the subject, submitted to {\it Zeitschrift f\"ur Physik} {\bf 46}
in 1944, has not appeared. Jordan referred to proof sheets which I
have not seen.} Jordan showed that Kaluza's field equations can be
simplified by the introduction of five homogeneous coordinates. Their
(homogeneous) transformations are isomorphic to the combined
coordinate and gauge transformations. No mention of his later
scalar-tensor theory is made in this note \cite{Jordan1945}. This was
made up in 1946 in a brief note in which the fifteenth field variable
of Kaluza's theory was identified by Jordan with the function to
replace the gravitational constant $\kappa$ \cite{Jordan1946}. In
projective relativity with projective coordinates $X^{\alpha}$, this
is expressed by by setting $J := g_{\alpha
  \beta}X^{\alpha}X^{\beta}~(\alpha, \beta = 0, 1,..,4$ with $x^{0}$ 
being the time coordinate) equal to $\frac{2\kappa}{c^2}$
\cite{JorMue1947}. In the article that followed, G. Ludwig (1918-2007) displayed
the relationship between variational principles in $^{(5)}R$ and
$^{(4)}R$ \cite{Ludwig1947}. The field equations for the gravitational field
$g_{ij}$, the electromagnetic 4-potential $ A_k=g_{4k}$, and the
$g_{44}$-variable $\kappa$ in space-time were then given by Jordan and
M\"uller\footnote{Contrary to common usage, Jordan \& M\"uller denoted
  the Ricci tensor in space-time by $G_{ik}$. The covariant derivative
  refers to $g_{ij}$. Equations (11) and (12) of  \cite{JorMue1947}
  corresponding to (\ref{Jordeq2}) and (\ref{Jordeq3}) here contain
  each a misprint. Both were corrected in \cite{Jordan1948}.} to 
be: \begin{eqnarray}G_{ik} + \frac{\kappa}{c^2}  
  F_i^{~s}F_{ks}= -\frac{1}{2\kappa} (\nabla_k\nabla_i \kappa -
    \frac{1}{2\kappa} \nabla_i\kappa~\nabla_k \kappa)~, \label{Jordeq1}\\
    \kappa \nabla_s F^{sj}=- \frac{3}{2}\nabla_s\kappa~
    F^{sj}~, \label{Jordeq2} \\ G = -  
    \frac{\kappa}{2c^2} F_{rs}F^{rs} +\frac{1}{2\kappa} g^{rs}
    \nabla_r\kappa~\nabla_s \kappa -
    \frac{1}{\kappa}g^{rs}\nabla_r\nabla_s
    \kappa~. \label{Jordeq3} \end{eqnarray} 
In January 1948, P. G. Bergmann (1915-2002) reported that work on a theory 
with a fifteenth field variable had been going on in
Princeton: \begin{quote}``Professor Einstein and the present author  
  had worked on that same idea several years earlier, but had finally
  rejected it and not published the abortive event''
  (\cite{Bergmann1948}, p. 255).\end{quote} It may be that, at the
time, they just did not have an idea for a physical interpretation
like the one suggested by P. Jordan. Although there were reasons for
studying the theory further, Bergmann pointed out that there is an
unwanted abundance in the theory: too many constructive
possibilities for a Lagrangian.\footnote{Actually, this was known by
  Jordan and his collaborators, cf. below.} Nonetheless, in his
subsequent paper on ``five-dimensional cosmology'', P. Jordan first
stuck to the simplest Lagrangian, i.e. to the Ricci scalar in {\em
  five} dimensions \cite{Jordan1948}:  
\begin{equation} ^{(5)}R =^{(4)}R + \frac{1}{4}J F_{rs}F^{rs} +
  \frac{2}{\sqrt{J} \sqrt{-g}}\partial_r(\sqrt{-g}g^{rs}\frac{\partial
  \sqrt{J}}{\partial x_{s}}). \label{5scal} \end{equation}

\subsection{Conceptional and formal developments}
\label{subsection: concdevelop}
However, his co-workers G\"unther Ludwig and Claus M\"uller in 1948
generalized the Lagrangian to \cite{LudMue1948b}:\begin{equation} J~ [^{(4)}R -
  \frac{J}{2} F_{rs}F^{rs} - (\lambda + \frac{1}{2}) \frac{\nabla_{r}J
    \nabla^{r}J}{J^2}] ~. \label{LudMueLag} \end{equation} Jordan had
sent the proof sheets of this paper to W. Pauli already on 14
Dec. 1947. Pauli pointed out that ``your theory is completely
equivalent with Kaluza's, if there is set $\frac{\partial}{\partial
  x^{5}}=0$, but $g_{55}$ is left arbitrary.'' He also found that
``the additional term makes the theory less inevitable''
(\cite{Pauli1993}, p. 510). Already in 1952, in the first edition of his book, 
Jordan acknowledged the work of Ludwig and M\"uller and exchanged
$^{(5)}R $ for the expression: \begin{equation} 
  \kappa^{\eta}~ [^{(4)}R - \zeta \frac{\nabla_{r}\kappa
    \nabla^{r}\kappa}{\kappa^2} -\frac{\kappa}{2c^{2}}F_{rs}F^{rs}]
  ~, \label{JorLag} \end{equation} where $F_{ij}$ is the
electromagnetic field, $\eta, \zeta (= \lambda + 
\frac{1}{2})$ dimensionless constants, and $J$ was replaced by
$\kappa$. Ludwig and M\"uller had taken $\eta= 1$.\footnote{In the
  1st edition of his book, Jordan not only gave (\ref{LudMueLag})  
and (\ref{JorLag}), but also $ J ^{\alpha}~[^{(4)}R - \lambda
\frac{\nabla_{r}J \nabla^{r}J}{J^2}]~ $ with numerical parameters
$\alpha, \lambda$.} In fact, in a further paper, Ludwig and M\"uller
had included the {\em matter}-Lagrangian for a perfect fluid and
obtained particular exact homogeneous and isotropic solutions for the
equations of state $p=0$ and $p=1/3 ~\rho$ \footnote{$p$ is the
  fluid's pressure and $\rho$ its density.}~\cite{LudMue1948a},
\cite{LudMue1948b}. One year later, Ludwig considered the general
Lagrangian with three free functions $U,V,W$:\begin{equation} {\cal L}=
  \sqrt{-g}~ [U(J) (^{(4)}R+ W(J) J_{,r} J_{,s}g^{rs}) +
  V(J)] ~ (r, s = 1, 2, .., 4) \label{LudLag} \end{equation} and
derived the corresponding field equations (\cite{Ludwig1949}, pp. 550, 552,
eq. (33)). About 20 years later, Bergmann \cite{Berg1968}, and Wagoner
\cite{Wagon1970} presented the same Lagrangian\footnote{with functions
  $h(\phi), l(\phi), \lambda(\phi)$ in Wagoner's paper replacing U, W,
  and V. Bergmann had four free functions $f_i(\phi), (i=1, 2, 3)$ and  
  $M(\phi)$ because he introduced the Maxwell Lagrangian separately
  instead of the matter Lagrangian.} (\ref{LudLag}); the subsequent
theory runs under the name ``Bergmann-Wagoner theory''
(\cite{Will1993}, pp. 123, 125). Ludwig's paper is not referred to:
both authors seemingly were unaware of its existence.   

By going from $g^{ij}$ to $ \psi^{-2}(\phi) g^{ij} $ and replacing
$g_{ij}$ by $h(\phi) g_{ij}$, the following form of the Lagrangian
with only two free functions was reached by Wagoner: \begin{equation}
  {\cal L}= \sqrt{-g}~ [^{(4)}R - n g^{rs} \frac{\partial
    \phi}{\partial x^r} \frac{\partial 
      \phi}{\partial x^s} + 2\lambda (\phi)] +  {\cal
      L}_{matter}(\psi^{2} g_{ij},..), ~(n=\pm 1)~. \end{equation} 
Ludwig seemingly had not been satisfied with the general Lagrangian
(\ref{LudLag}): it had not contained a matter term. So, in his book,
he presented the Lagrangian density as:\begin{eqnarray} J^{1/2}
  U(J)[(^{(4)}R+ \frac{1}{2} J F_{rs} F^{rs} + J^{-1}
  g^{rs}\nabla_r\nabla_s J + (W(J) - \frac{1}{2}J^{-2})  g^{rs} 
    \nabla_r J~\nabla_s J\\ + V(J) + U^{-1} L_{matter} \nonumber  ]\end{eqnarray}

Although for some time Jordan still clung to 5-dimensional theory, in
their papers and in the book summarizing them he and his coworkers
severed  the ``extended theory of gravitation'' from Kaluza's theory
and from the ``unitary field theory''-approach of the French group
around A. Lichnerowicz and Y. Thiry. G. Ludwig, C. M\"uller and
K. Just continued their research by treating the theory as a theory 
alternative to general relativity. The period August 1954 to
February 1956 was most fruitful for K. Just\footnote{Kurt Just, Professor 
Emeritus of Physics, University of Arizona, Tucson.}: he published 10 papers on
Jordan's gravitational theory plus one with G. Ludwig. Apart from
discussing possible Lagrangians and field equations for the theory, he
applied it to relevant problems like planetary motion, cosmology, the
torsion balance of E\"otv\"os, and to more specific questions as were
the Hertzian dipole or superpotentials. Aware of the ambiguities in the
choice of the Lagrangian clearly shown by the different papers of the 
group, he integrated them into the expression for the Lagrangian: 
\begin{equation} \kappa^{c}~ 
  ^{(4)}R - s \kappa^{c-2} \partial_{r} \kappa \partial_{s}\kappa
  g^{rs} +   \kappa^{c+1} L_{matter} ~, \label{JuLag3}\end{equation}
where $c$ replaced Jordan's parameter $\eta$ \cite{Jordan1955}. Jordan
had $c=1, s= \zeta$ (\cite{Jordan1952}, p. 140, eq. (7)), K. Just had
suggested $c=0$ \cite{Just1955c} but in the same year 1955 Ludwig \&
Just had also looked at $c = -1, s = 3/2$ (\cite{LudJu1955},
p. 474/75). In his application to the planetary system, more precisely
to perihelion motion, Just started from (\ref{JuLag3})
\cite{Just1956}. Just's proliferating publications probably did not help the
cause of Jordan's theory: it was prone to baffle readers. Jordan could
take notice of only one of Just's papers while proofreading the 2nd
edition of his book \cite{Just1954}. 

\subsection{Mass generation in the universe?}
\label{subsection:massgen}
In the 1st edition of his book, as a consequence of the field
equations, Jordan had obtained the equation $(\kappa^2 T^{\alpha\beta})_{; \beta} =
0$ and interpreted it as invalidating conservation of matter ``in the
old style, although, as required in principle by relativity, a definite form of
conservation law must still exist'' (\cite{Jordan1952}, p. 143). A
certain reservation is expressed by him when he wrote on the same
page: ``Therefore, we reserve the right to accept the theoretical
results only in part if the above equations are applied to processes of
matter generation.'' Nevertheless, he took it seriously: Applied to a
single star this meant that ``the single mass of a star $M_{St}$ must
have the property that its multiplication with $\kappa^2$ must yield
something constant: $\kappa^2M_{St}$= const.'' (ibid., p. 183). Also
in his letter to Pauli of 9 June 1953, Jordan defended his view, not without a portion 
of imaginativeness: ``[..] The calculation of the augmentation or generation of mass 
supposedly is only of statistical importance; real generation of mass probably does
not happen at the places where star-masses exist but obtains abruptly
with the generation of new stars, by fusion of our cosmos with
`embryonic stars' '' (\cite{Pauli1999}, p. 178).\footnote{``Die dann
  zu berechnende   Massenvermehrung oder Massenerzeugung geht wohl nicht an den Stellen
vorhandener Sternmassen vor sich, sondern erfolgt durch schlagartige
Erzeugung neuer Sterne durch Verschmelzung unseres Kosmos mit
`embryonalen Sternen.' ''} 

In the second edition, following Eddington and Pauli, another
interpretation was introduced: ``What is conserved, we call mass or
energy'' (\cite{Jordan1955}, p. 171). Consequently, the matter  
tensor is not $T^{\alpha\beta}$ but $\kappa^2 T^{\alpha\beta}$. This
was criticised in a book review by O. Heckmann who argued that then the
meaning of $\kappa$ as gravitational coupling function again is questionable. 
He also found Jordan's ``cosmogonic'' consideration very speculative: 
``he abandons the ancient [..] assumption of the conservation of matter'' 
(\cite{Heck1956}, p. 282).  

\subsection{Criticism by Pauli and Fierz}
\label{subsection:criticism}
Jordan's theory received wider attention after his and G. Ludwig's books had 
been published in the early 1950s \cite{Jordan1952}, \cite{Ludwig1951}. In a 
letter to Jordan, Pauli criticized projective relativity as bringing
no progress with regard 
to Kaluza's theory and questioned Jordan's taking the five-dimensional
curvature scalar $^{(5)}R$ as his Lagrangian.  

One of those responding also to this book was M. Fierz in
Basel. Previous to the publication of his paper \cite{Fierz1956}, he had
corresponded with W. Pauli. In following an idea of Lichnerowicz
(\cite{Lichno1955}, p. 201), he interpreted the scalar function as
``permittivity of the vacuum'' (the electric constant) 
 $\epsilon_0 =1/\mu_0 = \kappa^2$ and pointed to a difficulty of coupling the
scalar field to the energy-momentum tensor of matter. In fact, for the
transition from the gravitational constant $\kappa$ to a scalar field
function $\phi$, it makes a difference whether the Einstein Lagrangian
is written as $^{(4)}R + \kappa L_{matter}$ or as $\kappa^{-1}~
^{(4)}R + L_{matter}$. This is one of the differences between Jordan's
and Brans-Dicke scalar-tensor theory. Fierz suggested to introduce
either point particles as matter or a (quantum-) Klein-Gordon field in
order to remove conformal invariance. He sent the first versions of his paper 
to Pauli and eventually received Pauli's placet; cf. the letter of W. P. to
M. F. of 2 June 1956 in (\cite{Pauli2001}, p. 578). 

In the 2nd edition
of his book, Jordan then discussed a difficulty of his theory pointed
out by W. Pauli in this context: instead of $g_{ik}$ equally well
$\phi(x)g_{ik}$ with arbitrary function $\phi$ could serve as a
metric. Jordan's theory is conformally invariant only in the case that
an electromagnetic field forms the matter tensor (particles with zero
rest-mass). A problem for the interpretation of mathematical objects
as physical variables results: by a suitable choice of the conformal
factor $\phi$, a ``constant'' gravitational coupling function could be
reached, again. In the 2nd edition of his book, Jordan introduced a
new section on ``Pauli's conform-transformations'' (\cite{Jordan1955},
\S 28). Both, Pauli and Fierz gave a low rating to Jordan's
theory\footnote{See letters of W. P. to  M. F. of 30. 9. 1955, p. 350 
  and 2. 3. 1956, p. 531; of M. F. to W. P. of 8/11. 3. 1956, p. 539
  in \cite{Pauli2001}.}  

In 1959, Jordan had fully accepted the criticism by Pauli and
Fierz: \begin{quote} ``If one goes along with the considerations
  communicated by Fierz - and it seems to me that they are imperative
  - then apparently Dirac's hypotheses II can no longer be upheld,
  because [..] none of the field equations derivable from a variational
  principle [..] permits a violation of mass
  conservation.''\footnote{Schlie{\ss}t man sich den von Fierz  
    mitgeteilten Erw\"agungen an - und es scheint mir, da{\ss} sie
    tats\"achlich zwingend sind - so scheint es zun\"achst, da{\ss}
    die Diracsche Hypothese II nicht mehr vertretbar ist, da weder [..]
    noch irgendwelche aus einem Variationsprinzip ableitbaren
    Feldgleichungen [..] eine Durchbrechung der Massenerhaltung
    zulassen'' \cite{Jordan1959}.} \end{quote} As to the special
choice of $c~(\eta) =-1$ in eq. (\ref{JuLag3}), his argument was that
otherwise, as had been noted  before by Lichnerowicz for  $c~(\eta) =+1$,
 the permittivity of the vacuum $\epsilon_{0} = \kappa^{1+\frac{1}{\eta}}$
would depend on $\kappa$.  

\section{Scherrer and elementary particles} 
Neither Pauli nor Fierz, lecturer in Basel, Switzerland, at the time,
payed attention to the fact that a mathematician at the university of Bern had
suggested a scalar-tensor theory already in 1941 {\em before} P. Jordan,
and without alluding to Kaluza's theory or to Pauli's projective
formulation. The short note, written in German, was not readily
accessible outside the German speaking countries and carried the
title ``Zur Theorie der Elementarteilchen'' (About the theory of elementary 
particles) \cite{Scherrer1941}. In this note, the mathematician Willy Scherrer 
(1894-1979) of Bern, Switzerland, introduced a scalar field $T(x^0, x^1, x^2, x^3)$ 
and a variational principle \begin{equation} \delta \int T  R \sqrt{-g} d^4x = 0~, 
\label{Scher1} \end{equation} with the constraint \begin{equation} 
\int T \sqrt{-g} d^4x = constant~, \end{equation} where $R$ is the curvature 
scalar of space-time. The constraint reminds us of the norming of a
Schr\"odinger  
wave function ($T \sim \psi \bar{\psi}$) and, in fact, Scherrer gave as his
motivation ``new Ans\"atze for a scalar relativistic wave
mechanics''.\footnote{``Veranlasst durch neue Ans\"atze zu einer
  skalaren relativistischen Wellenmechanik [..].''} In the same year,
Scherrer had tried to describe the interaction of two particles by
means of, as he said, ``relativistic Schr\"odinger equations''
\cite{Scherrer1941a}, \cite{Scherrer1941b}. He finally had come up
with a system of two coupled relativistic wave equations for two wave
functions $u, v$. Interestingly,  in his correspondence with W. Pauli
in connection  
with a discussion of Dirac's equation in his extended theory of gravitation, Jordan 
claimed that $\int \kappa^2 \psi \psi^{*}$ is time-independent. Pauli agreed
and gave as the commutation relations of ``second quantization'':
$\kappa^2 \{\psi_{\alpha} (x, t), \psi^{*}_{\beta} (x', t)  \}_{+}= \delta_{\alpha \beta}
\delta^{(3)}(x-x')$. But he added `` [..] everything goes as if $\Psi =
\kappa \psi$ is introduced as a new spinor and $\kappa$ is omitted''
(\cite{Pauli1999}, pp. 178, 180, 194.)  

In his first note, Scherrer proposed a generalization of his
Lagrangian in (\ref{Scher1}) to $ T[R + k (grad~ln~ T)^{2}]$ with a
constant $k$ and $(grad~ ln~ T)^{2}$ obviously abbreviating
$\frac{k}{T^{2}} g^{rs} \frac{\partial T}{\partial x^r} \frac{\partial
  T}{\partial x^s}$. This corresponds to Jordan's theory.\\ 

Six years later, Scherrer had seen a paper by Ludwig \& M\"uller
\cite{LudMue1948a} and probably hastened to secure priority for
himself by elaborating on his previous note. He now presented another
argument for the replacement of Einstein's equations (\cite{Scherrer1949}, p. 537): 
``In the case of vanishing matter, one should expect degeneracy of the metrical
structure from a theory which makes matter responsible not just for
deviations from geodesic motion but for the total metrical 
structure.'' \footnote{``Von einer Theorie, die die Materie nicht nur
  f\"ur die Abweichungen von der Tr\"agheitsbahn, sondern f\"ur die
  totale metrische Struktur verantwortlich macht, sollte man
  eigentlich erwarten, dass sie im Falle verschwindender Materie
  entartet.''} As a cure for this apparent deficiency, he suggested to
allocate to each place in the world an ``intensity'' $ \psi^2 $ such
that $\psi^2 \sqrt{-g}d^4x$ stands for the relative number of material
elements lying in the volume element $ \sqrt{-g}d^4x$. Thence, he interpreted 
his theory as an alternative to Einstein`s theory.

In place of the manifest $^{(4)}R \psi^2 \sqrt{-g}d^4x$, Scherrer
advocated the more general Lagrangian of his note of 1941
(\cite{Scherrer1941}, p. 238): \begin{equation} {\cal L} = [(^{(4)}R -2
  \Lambda)   \psi^2 + 4\omega g^{rs} \frac{\partial \psi}{\partial
    x^r} \frac{\partial \psi}{\partial
    x^s}]\sqrt{-g}~.\label{ScherLag}\end{equation} $\psi$ corresponds 
  to a scalar matter field.\footnote{Scherrer had also thought about a vector
    field as matter, but discarded the idea. Cf. (\cite{Scherrer1949}, p. 539).} In the paper, 
he showed that for $\Lambda = \omega = 0$ a static and spherically symmetric
  exact 2-parameter solution existed containing the Schwarzschild solution as a
  special case. He calculated its ``total energy'' $\int
  T_{0}^{0}\sqrt{-g}d^4x$ to be finite. The model-system being in
  ``complete equilibrium'' according to Scherrer, ``it logically first
  should be applied to elementary particles''\footnote{ ``[..] muss es
    sinngem\"ass in erster Linie f\"ur Elementarteilchen in Aussicht
    genommen werden.''} He gave two reasons for why quantization had
  not yet occurred in his theory: (1) The restriction to a static
  situation might be too restrictive and, (2) quantization possibly
  might first show up in a 2-body-problem. Scherrer´s static, spherically
symmetric solution preceded the correponding solution by Heckmann et al. 
\cite{HeJoFri1951} by one year. 

Scherrer advised a student, K. Fink, to continue his approach. But Fink
looked at the Lagrangian density $ {\cal L} = (R
+ 2\omega g^{rs} \frac{\partial \psi}{\partial x^r} \frac{\partial
  \psi}{\partial x^s})\sqrt{-g}~  $\cite{Fink1951}. In it the coupling of 
the scalar matter field $\psi$ to the curvature scalar is missing! The
ensuing field equations correspond to Einstein's equations for a
massless scalar field or, equivalently to those following from
Jordan's Lagrangian if the parameter $\eta = 0$ (\cite{Jordan1952}
p. 140). Exact solutions in the static, spherically symmetric case and
for a homogeneous and isotropic cosmological model in these differing theories 
were submitted in 1951 almost on the same day by Fink ($\eta=0$) and Heckmann, 
Jordan \& Fricke ($\eta=1$) \cite{HeJoFri1951}.

\section{Thiry and mathematical physics} 
 
In January 1950, Yves Thiry 
submitted a thesis to the faculty of science of
Paris University with the title ``Mathematical study of the equations
of a unitary theory with fifteen field variables''. He exuberantly
thanked his ``master and friend Lichnerowicz'' who obviously had
initiated the work. Unfortunately, ``Jordan and his school'' had
``obtained almost at the same time like us the equations which we will
give in Chapter II. We had no knowledge about this except at a very
late stage, and it is only recently that we could correspond with
Jordan. He was so friendly as to send us his publications which we
could not have procured otherwise.'' (\cite{Thiry1951}, p. 6) 
\footnote{``[..] obtenu \`a peu pr\`es en m\^eme temps que nous
  les \'equations que nous donnons au Chapitre II. Nous n'avons eu
  connaissance de ce fait  que fort tard et ce n'est que r\'ecemmant
  que nous avons pu correspondre avec Jordan, qui a eu l'amabilit\'e
  de nous envoyer ses publications qu'il \'etait alors impossible de
  se procurer autrement.''} (\cite{Thiry1951}, p. 6) In fact, it was
A. Lichnerowicz (1915-1998), then at the University of Strasbourg, who had written
to W. Pauli and asked for ``Jordan's original paper'' (cf letter of
W. Pauli to P. Jordan of 23. 3. 1948 in \cite{Pauli1993},
p. 516). According to Pauli: \begin{quote} ``Lichnerowicz is a pure
  mathematician who is occupied with the integration of Einstein's
  field equations. One of his students, Ives Thiry now has looked into
  the (not mutilated) Kaluza-theory (with $g_{55}$) and, so I believe,
  has simplified very much the calculational
  technique.''\footnote{``Lichnerowicz ist ein reiner 
  Mathematiker, der sich mit der Integration der Einsteinschen
  Feldgleichungen befasst. Einer seiner Sch\"uler, Ives Thiry hat
  sich nun mit der (unverst\"ummelten) Kaluza-Theorie (mit $g_{55}$)
  besch\"aftigt und, glaube ich, die Rechentechnik sehr
  vereinfacht.''}\end{quote} Since 1947/48, first together with
A. Lichnerowicz, Thiry had published short notes about Kaluza's
theory in Comptes Rendus. In the first note of February 1947 on
variational calculus, after presenting a certain mathematical formula they
stated: ``Therefore, Kaluza's theory presents itself as an immediate
application of formula (5) which even may lead to an extension of
this theory to a theory with $g_{00}\neq 0$'' (\cite{LichnoThir1947},
p. 531). The 15 field equations then were given by Thiry in January 1948
  \cite{Thiry1948a}.\footnote{In the third paper, the global problem
    whether regular solutions exist was dealt with \cite{Thiry1948b}.} On 19 
January 1948 \footnote{The call number of the Einstein Collected Papers (ECP) 
is 16-312.00.}, he sent this second note in Comptes Rendus  to Albert Einstein 
\cite{Thiry1948a}. In his publication preceding Lichnerowicz' letter to Pauli, 
Thiry had not yet given a physical interpretation of the scalar field \cite{Thiry1948a}. 
Around that time, the interest in 5-dimensional relativity seems to have risen; we 
already have met P. G. Bergmann's paper of 1948 \cite{Bergmann1948}. 
\footnote{Bergmann's paper appeared  only on January 1, 1948 although it had 
been submitted on August 30, 1946. Thus he could not yet have reacted to Thiry's
correspondence with Einstein.} C. V. Jonsson, a student of O. Klein in
Stockholm, also wrote a long paper about the theory's field equations. He 
included the scalar field and dropped the cylinder condition.\footnote{I have 
not been able to verify that Jonsson's field equations for the case of the cylinder 
condition agree with Thiry's equations.} He then  
quantized the free field in linear approximation \cite{Jons1951}.  
Thiry's interest in Kaluza's theory was of mathematical nature: \begin{quote} 
``As to unitary field theories, it seems that their mathematical study has been 
quite neglected [..]. We thought it useful to try a systematic
mathematical study of a unitary field theory, and to find out whether
such a theory is able to present the same coherence like general
relativity.''  (\cite{Thiry1951}, p. 3) \footnote{``Quant aux
  theories unitaires, il semble que leur \'etude math\'ematique ait
  \'et\'e relativement neglig\'ee [..] Il nous a paru utile de tenter
  une \'etude math\'ematique syst\'ematique d'une th\'eorie unitaire
  et de voir si une telle th\'eorie est susceptible de pr\'esenter la
  m\^eme coh\'erence que la th\'eorie de la Relativit\'e
  g\'en\'erale.'' }\end{quote} 
In three chapters, Thiry's thesis laid out the conceptual background of
a 5-dimensional theory, the setting up and study of the field equations by 
help of Cartan's differential calculus, and results on regular solutions of 
the theory's field equations. Unlike in the approach by Einstein \&
Bergmann, the cylinder condition $g_{\alpha \beta ,4}=0$ is upheld,
throughout. Here he used an argument from physics: no physical
phenomena furnished evidence for the existence of a fifth dimension 
(\cite{Thiry1951}, p. 39). His access to 5-dimensional space used the
fact that the equations of motion of charged particles are geodesics
in Finsler geometry; for each value of $\frac{e}{m}$ (charge over
mass) another Finsler space is needed with the
metric:\begin{equation}ds = \sqrt{g_{jk}dx^jdx^k} + \frac{e}{m}A_l
  dx^l~.\label{Finsmet1}\end{equation} He then showed that a
5-dimensional Riemannian space could house all these
geodesics.\footnote{In nuce, this idea can already be found in his 
  paper with Lichnerowicz (\cite{LichnoThir1947}, p. 531).} 
\begin{quote} The introduction of a fifth coordinate [..] thus shall
  justify itself by the fact that it imparts the role of geodesics to the
  trajectories of charged particles which they lost in space-time
  [..]''\footnote{``L'introduction d'une cinqui\`eme
    coordonn\'ee,[..] se justifiera donc par le fait qu'elle
    confi\`ere aux trajectories des particules \'electris\'ees le
    r\^ole de g\'eodesiques qu'elles perdaient dans l'espace-temps
    [..]''}\end{quote} In the third chapter, Thiry aimed at showing
that his unitary field theory possessed the same mathematical
coherence with regard to its global aspects. By partially using
methods developed by Lichnerowicz, he proved theorems on the global
regularity of solutions. These results are of a different nature 
than what Jordan had achieved; they are new and mathematically rigorous.

Yet Jordan was not in a hurry to read Thiry's thesis; he wrote to Pauli: 
\begin{quote}''By the way, in his th\`ese published in 1951, Thiry has
  studied systematically and extensively the theory with variable
  gravitational constant; [..] I received it only after my book appeared
  and, at present, I have not read it very closely. It thus is not
  really clear to me whether it contains interesting novelties.''
  (\cite{Pauli1996}, p. 799/800)\footnote{''Thiry hat ja
    \"ubrigens in seiner 1951 ver\"offentlichten Th\`ese die Theorie
    mit variabler Gravitationskonstante systematisch und ausf\"uhrlich
    studiert;[..] Ich habe es erst nach dem Erscheinen meines Buches
    von ihm bekommen und augenblicklich noch nicht sehr genau
    gelesen. Ich wei{\ss} also auch noch nicht recht, ob interessante
    Neuigkeiten darin stehen.''} \end{quote} To Pauli, Thiry's global theorems
might not have been ``interesting novelties'', because in his
corresponding paper with Einstein on the non-regularity of solutions,
the proof had been independent of the dimension of space
\cite{EinPaul1943}. Pauli, at first, also did not read Thiry's th\`ese, but 
responded arrogantly:\begin{quote} ``The Th\`eses by
  Thiry are laying on my desk; however they are so appalingly thick (do
  not contain a reasonable abstract) such that it is so much simpler
  to not open the book and reflect about what must be inside.''
  (W. Pauli to P. Jordan 8. 6. 1953, \cite{Pauli1999},
  p. 176)\footnote{''Die Th\`eses von Thiry liegen auf meinem
    Tisch; sie sind aber so entsetzlich dick (haben auch keine
    vern\"unftige Zusammenfassung), da{\ss} es so viel einfacher ist,
    das Buch nicht aufzumachen und sich zu \"uberlegen, was darin
    stehen muss.'' - Thiry's thesis comprises 122 pages.}\end{quote} Somewhat 
later, Pauli corrected himself and wrote to Jordan that in the
preparation for a course ``he nevertheless had read around in Thiry's
Th\`ese'' (W. Pauli to P. Jordan 3. 2. 1954, \cite{Pauli1999},
p. 442). Note that neither of these two eminent theoretical physicists
discussed Thiry's paper as regards its valuable content. 

As to the field equations corresponding to eqs. (\ref{Jordeq1}) to
(\ref{Jordeq3}), Thiry had calculated them in great detail with
Cartan's rep\`ere mobile for both a euclidean or Lorentz 
metric of $V_5$, and also with a 5-dimensional matter tensor of the form of dust
$\rho u^{\alpha} u^{\beta}$. From the 15th equation, he even had
obtained ``a new physical effect'': uncharged dust-matter could
generate  an electromagnetic field  (\cite{Thiry1951}, p. 79,  
footnote (1)). He linked this effect to Blackett's search for the
magnetic field of a gravitating rotating body. Lichnerowicz and his doctoral
student Francoise Hennequin discussed consequences of the projection  
of the matter tensor into space-time: a charged perfect fluid could be
described. This included the interpretation of $\kappa =
\frac{G}{\phi}$ as ``a gravitational factor'' in front of the matter  
tensor in space-time reducing to the gravitational constant for $\phi=
1$ by Lichnerowicz (\cite{Lichno1955}, p. 202). Thiry himself kept
Jordan's original choice $^{(5)}R$ for the Lagrangian.

\section{Princeton: R. Dicke and Mach's principle}
Since the mid 50s, Robert H. Dicke (1916-1997) had been thinking about Dirac`s large number hypothesis, the equivalence principle, and Mach's principle \cite{Dicke1959a}, \cite{Dicke1959b}. According to Sciama,
an expression for this principle is given by  $G V=-c^{2}$, where V is the gravitational potential of the universe and $G$ the gravitational constant. Qualitatively , thus $GM/Rc^{2}\sim 1$, whith $M$ the mass of the observable universe, $R$ the radius of the boundary of the observable universe, and $c$ the vacuum velocity of light \cite{Sciama1953}.

In another paper, Dicke may have taken up the idea of Fierz when he
wrote: ``that a space variation in the polarizability of the vacuum
will lead to a number of results familiar as typical gravitational
effects.'' In this context, he investigated ``a form which a theory of
gravitation may take when the principle of equivalence is satisfied in
a weakened form only'' (\cite{Dicke1957a}, p. 363). He went on:
``Jordan has previously considered a similar problem and Fierz has
made a critical  analysis of Jordan's theory.'' References to Jordan's
book (2nd edition 1955) and the paper of Fierz are
given.\footnote{According to Dicke (p. 364): ``The fact that many of
  the properties of gravitation can be accounted for in terms of an
  interaction with a polarizable medium is an old idea which has
  recurred from time to time.'' He then cited a paper of H. A. Wilson
  of 1921.} The theory was to accept Mach's principle, the
cosmological principle, and general covariance. A theory for a scalar
field $\epsilon$ in a flat background is suggested having the   
Lagrangian: $L=\frac{1}{k}h^{ij}\epsilon_{,i}\epsilon_{,j}$ with
$h^{ij}$ being the reciprocal of the symmetric tensor $g_{ij}$
describing the vacuum with $ g_{11}= g_{22}= g_{33}= -\epsilon;~
g_{44}=1/\epsilon$. The quantity $ \epsilon$ is taken as the electric
specific inductive capacity of the vacuum. This looks rather different
from the motivation of Scherrer or Jordan for their scalar-tensor
theories theories. It turns out that, in Dicke's approach,
asymptotically, the scalar field is proportional to the inverse of the
ratio of the gravitational to the electromagnetic interaction between
two elementary particles on an atomic scale. Thus, a connection to
Dirac's hypothesis is established (\cite{Dicke1957a}, p. 375). \\ 

What now is called Brans-Dicke scalar-tensor theory then was published
in 1961 by C. H. Brans and R. H. Dicke \cite{BraDick1961} and in a
subsequent paper by Brans \cite{Brans1962}. In the joint paper, no
reference was given to Brans' dissertation submitted to the Princeton
faculty in May 1961 by Brans \cite{Brans1961}. In the first
publication, Dicke's previous paper of 1957 was also not
mentioned. The authors were at pains to distance their theory from
Jordan's: ``There is a formal connection between this theory and that
of Jordan, but there are differences and the physical interpretation
is quite different'' (\cite{BraDick1961}, p. 928). As a second
reference besides Jordan's book of 1955, his paper of 1959
\cite{Jordan1959} was cited with the comment: ``In this second
reference, Jordan has taken cognizance of the objections of Fierz [..]
and has written his variational principle in a form which differs in
only two respects from that expressed in Eq. (16).'' \footnote{The
  reference to Eq. (16) is a typo; Eq. (6) of   \cite{BraDick1961}
  must be meant.} The name ``Brans-Dicke theory'' was used by Dicke
right away (\cite{Dicke1962a}, p. 2167), \cite{Dicke1962b}, p. 656,
caption of figure 1).     

The Lagrangian presented by Brans and Dicke is: 
\begin{equation} {\cal L} = (R  \phi -\frac{\omega}{\phi} g^{rs}
  \frac{\partial \phi}{\partial x^r} \frac{\partial \phi}{\partial
    x^s})~\sqrt{-g}~. \label{BraDiLag} \end{equation} This is exactly
Scherrer's expression (\ref{ScherLag}) published in 1941 and 1949 if
$\phi= \psi^2$, the cosmological constant $\Lambda=0$, and Scherrer's
constant $\omega$ is replaced by the constant $-\omega$ in the
publication of Brans and Dicke.\footnote{However, the definitions of
  the curvature tensor by Scherrer and by Dicke-Brans differ by a
  minus sign. Thus, in the results the symbol $\omega$ stays the
  same. Cf. (\ref{Schereq}) below.} The field equations obtained from
(\ref{ScherLag}) are: \begin{equation}  
R_{ij}-\frac{1}{2}g_{ij} R = \frac{8\pi}{c^{4}} \phi^{-1}T_{ij} +
\frac{\omega}{\phi^{2}} (\phi_{,i}       \phi_{,j} - \frac{1}{2}g_{ij}
\phi_{,r}\phi_{,s} g^{rs}) + \phi^{-1} (\phi_{,i;j} - g_{ij} \square
\phi)~.       \label{BD1}  \end{equation} This equation had been given
already in 1948 by Ludwig and M\"uller \cite{LudMue1948a},
\cite{LudMue1948b}. The scalar wave equation following from
(\ref{BraDiLag}) is: 
\begin{equation}2\omega\phi^{-1} \square \phi -\omega
  \phi^{-2}\phi^{,i}\phi_{,i} + R = 0~; \label{BD2}\end{equation} it
had already been derived in 1948 by Ludwig and M\"uller
\cite{LudMue1948a}, \cite{LudMue1948b}. Now, the advantage of the
choice $\kappa = \phi^{-1}$ in Brans-Dicke theory is that the
field equation for $\phi$, after manipulation of eqs. (\ref{BD1}),
(\ref{BD2}), becomes the wave equation: \begin{equation}\square \phi =
  \frac{8\pi T}{2\omega + 3}~, \end{equation} where $T$ is the trace
of the matter tensor, and $\square \phi = g^{rs} \nabla_r \nabla_s
\phi$ the covariant d'Alembertian. Sciama's expression for Mach’s
principle  is an immediate consequence in the static case. It is to be noted that
Scherrer had \footnote{His eq. (2.19) on p. 542;
  cf. \cite{Scherrer1949}.} \begin{equation}(3+2\omega) \square \chi
  - 2\Lambda \chi = 0 ~.\label{Schereq}\end{equation}   

The two differing points mentioned above were: (1) Jordan set $\kappa
= \phi$, while Dicke and Brans chose $\kappa = \phi^{-1}$; (2) ``[..]
as a result of its outgrowth from his five-dimensional theory, Jordan
has limited his matter variables to those of the electromagnetic
field.'' In view of Jordan's discussion of the matter tensor in
connection with its non-conservation, and the presentation of
cosmological solutions for an ideal fluid in \S 30 of his book
(\cite{Jordan1955}, pp. 186-196), this second point looks rather contrived. 
The publication by Brans and Dicke was a first short summary of the
dissertation of C. H. Brans. It included the weak-field equations, a
static, spherically symmetric exact solution applied to the
calculation of the perihelion shift and light deflection deviating
from Einstein's theory by the factors $\frac{3+2\omega}{4+2\omega}$ 
and $\frac{3\omega + 4}{3\omega + 6}$, respectively.  These values had
been calculated before by Heckmann et al. (\cite{HeJoFri1951},
pp. 139, 141).\footnote{For the parameters in both papers we have
  $\alpha_{0} = -\frac{1}{2+\omega}$.} Also, a derivation of Mach's
principle in the form $GM/Rc^{2}\sim 1$, and a discussion of
homogeneous and isotropic cosmological models were given. The
follow-up by Brans alone contained a thorough investigations of the
relationship of a locally measured gravitational constant and ``the
structure of the universe'' as well a discussion of the boundary
conditions for the scalar field, its consequences for a spherically
symmetric mass distribution, and conservation laws
(cf. \cite{Brans1962}, sections II-IV).  

All this had been worked out in detail in Brans' dissertation. In two
chapters  of it (VI C, p. 37-38 and VIII A-G, pp. 57-68), he discussed
Jordan's work as presented in the 2nd edition of his book, in a later
paper of  1959 \cite{Jordan1959}, and also publications of K. Just 
\cite{Just1955a}, \cite{Just1955b}, \cite{Just1956} and G. Ludwig
\cite{LudJu1955} from 1955 and 1956. In these papers, Jordan's theory
had been applied to planetary motion, precession of the perihelion and
to cosmology. Also, his theory of the generation of matter had been
discussed. In his thesis, Brans made it very clear that one of the
main incentives for Dicke's and his work was to keep the weak
principle of equivalence and to leave unchanged `` the entire theory
of matter and electromagnetism'' plus  the ``Coulomb determination of
inertial mass used in general relativity'' (\cite{Brans1961},
p. 36). However, Dicke's ``strong'' principle of equivalence was
violated.\footnote{Brans defined Dicke's strong equivalence principle:
  ``as the assertion that in the absence of non-inertial and
  non-gravitational forces, the numerical content of experiments  
performed in a locally almost flat physical coordinate system is
independent of any characteristics of the mass distribution in the
rest of the universe'' (\cite{Brans1961}, p. 26).} The spherically
symmetric vacuum solution of (\ref{BD1}), (\ref{BD2}) by Heckmann and
Fricke, discussed in \S 29 of Jordan's book was presented and compared
to the exact solutions Brans had obtained himself in isotropic
coordinates. Thus, while Brans carefully examined Jordan's theory as
presented in and after 1955, he and Dicke were silent on the earlier
papers of the Hamburg group and on the contributions of the Paris
group. This may be due to their belief that it be unnecessary to cite
earlier references already listed in the second edition of Jordan's
book. Nevertheless, this made appear the work of Dicke and Brans
prompter in time concerning the time-scale of publications. They
rightly felt no need to cite the paper by Jonsson of 1951
\cite{Jons1951}, written in English and refered to in Jordan's book
because Jonsson's paper dealt with 5-dimensional space and did not
suggest another scalar-tensor theory.    

\section{Stockholm's H\"ogskola: Jonsson and Klein}
\label{section:jonsson}
C. W. Jonsson's knowledge of the literature in the field in which he
set out to work must have been limited: although starting from
5-dimensional theory, he did not mention Kaluza's name. This is the
more strange as his work was done under supervision by Oskar Klein (1894-1977)
in Stockholm who had given Kaluza's theory a new twist in the 20s. Klein
had seen 5-dimensional space essentially as a four dimensional one
with a small periodical strip or tube in the additional spacelike
dimension affixed. The 4-dimensional metric then is periodic in the
additional coordinate. In 1946, O. Klein had tried to geometrize the
meson field via Kaluza's 5-dimensional theory and allowed all field
quantitied to be periodic functions in the additional coordinate
\cite{Klein1947}. In view of the fact that Jonsson derived the
field equations {\em without prescribing the cylinder condition},
i.e.,  by allowing the components of the 5-dimensional metric to be
dependent on the fifth coordinate, he also could have cited
Einstein and Bergmann. In their new approach in 1938, they had
claimed to ascribe ``physical reality to the fifth dimension
whereas in Kaluza's theory this fifth dimension was introduced only in
order to obtain new components of the metric tensor representing the
electromagnetic field'' (\cite{EinBerg1938}, p. 683). That Klein
advised a student to extend Kaluza's theory by allowing the additional fifth
coordinate in all field variables, may have been a consequence of
Pauli's violent reaction to Klein's talk given in Princeton in winter 1949/50. 
Klein then hat insisted on keeping the cylinder condition
(cf. \cite{Pauli1996}, p. XXV).

Moreover, Jordan and his co-workers remained uncited. Certainly, Jordan's book 
had appeared only in 1952, but he and his collaborators had published papers 
before in Annalen der Physik and Zeitschrift f\"ur Naturforschung both of which 
might have been available in Stockholm at the time. \footnote{Sweden
  was a neutral country during  the second world war such that the
  flow of scientific communication with  Germany should not have been
  hampered; this includes the   period directly after the war. On the
  contrary, according to W. Pauli, during the war it was impossible to
  send printed matter from the USA to Sweden.} Also, G. Ludwig's book
had appeared in 1951, which contained ``the results of some papers of
P. Jordan, G. M\"uller and the author as well as some not yet
published results'' (\cite{Ludwig1951}, p. 3). Thiry's thesis was
mentioned by Jonsson although it had just been published at the
beginning of 1951. Jordan himself felt encouraged by Jonsson's
paper. In it, as in Thiry's approach, the same extension of Kaluza's
theory had been performed which he had suggested; hence he no longer
could be seen as an outsider (\cite{Jordan1955}, p. 161). There is a
correspondence between Pauli and O. Klein about Jordan's theory only
later in 1953 (\cite{Pauli1999}, p. 209).\\ 

Whatever the reasons for Jonsson’s negligent way of citing may have been,
his motivations were quite different than those used in the other approaches 
described above: he was not interested in scalar-tensor theory as a
theory of gravitation but in its field quantization in linear
approximation. Hence he set $\gamma_{\mu \nu} =
\delta_{\mu \nu} + \epsilon \phi_{\mu \nu} $ with $| \phi_{\mu \nu}|<<
1~; (\mu, \nu = 0, 1, 2,..4)$ The conditions $\frac{\partial \phi_{\mu
    4}}{\partial x^{4}} = 0$ now were added. \footnote{$ x^{4}$ is  
the fourth space-coordinate; $x^{0}$ denotes the time-coordinate.}
$\phi_{rs}$ with $ r, s, =0, 1, 2, 3$ was assumed to be periodic in
$x^{4}$ and expanded in a Fourier series. With the scalar field set
constant, the quantization of the free field of spin $2\hbar$ was
then carried through in Fock-space. C. V. Jonsson seems to have left
no further traces in theoretical physics.  

\section{Problems in communication}
In his personal memories, Carl Brans stated that during the writing
of his thesis: ``[..] I discovered the work of Jordan et al. on this
topic and almost quit writing. However, I was encouraged to  
continue, giving what I hope is sufficient and appropriate credit to
Jordan and his group.'' (\cite{Brans2008}, p. 5) He does not say how
he discovered Jordan. In view of the references to Jordan's book given by his 
thesis adviser Dicke already in two publications of 1957
(\cite{Dicke1957a}, p. 363; \cite{Dicke1957b}, p. 356), it seems that
the discovery cannot have been of an abrupt nature. At the same time
at which Brans wrote up his thesis, the Princeton postdoc Dieter Brill
\footnote{Dieter Brill, professor at the University of Maryland.}
spent the year 1960/61 in Hamburg at the Physikalisches Staatsinstitut
as an independent Flick exchange fellow. There has been some contact
between  Brill and Dicke in the preparation of the Enrico Fermi Summer
School of 1962, where both gave lectures with Brill reviewing Jordan's
theory \cite{Brill1962}. His detailed presentation of results by
Jordan and some of his associates (among others from K. Just's
Habilitationsschrift) was the first one for English speaking
relativists. However, there has been no direct interaction between
D. Brill and C. Brans concerning Jordan.\footnote{Private
  communication of 8 April 2012 by D. Brill to the author.} It also is
unlikely that there has been a correspondence between Brans' thesis
adviser R. Dicke and P. Jordan as in the case of Lichnerowicz and Thiry.  

D. Brill speaks German perfectly; Carl Brans also can read German.
Moreover, Peter Bergmann who had not cited the paper of G. Ludwig was
of German origin and thus could read the language fluently. In
Princeton, Valentin Bargmann whom Dicke had thanked in his paper on a
special relativistic treatment of gravity, although of Russian origin,
had worked long enough with Wentzel, Pauli, Bergmann and Einstein to
be proficient in German. Thus, with regard to the papers of Jordan and
his associates, language problems should not have played a major role in 
Princeton. With French, it might have been different. Mme. Tonnelat's book on
unified field theory of 1955 \cite{Tonnel1955} was translated into
English only in 1966. Lichnerowicz' book on relativistic theories of
gravitation of the same year \cite{Lichno1955} in which his and Thiry`s results 
were presented and Tonnelat`s other books remain untranslated until today.\\  

What is called now ``Community of Scholars'' at the Princeton
Institute for Advanced Study (PIAS) then seems to have been a real
in-group. It included W. Pauli who had spent the summer term of 1954
there with previous stays in 1935-36, 1940-1946, and 1949-50;
V. Bargmann had spent terms at PIAS in 1938-46 and 1954-55; M. Fierz
stayed there during the fall term of 1950, Oskar Klein spent a term at
PIAS in 1949-50 giving atomic physics and relativity theory as his
interests. Neither someone from the Hamburg group (Jordan, Ludwig,
Just) nor from the French groups (Lichnerowicz, Thiry, Tonnelat) has
resided at the Institute for Advanced Studies in Princeton. Two
scientists could have acted as go-between with regard to Paris and
Princeton, C\'ecile DeWitt-Morette, at PIAS in 1948-1950, and Yvonne
Choquet-Bruhat in 1951-52. But they had shown no special interest in
gravity; DeWitt-Morette while working in theoretical physics was
interested in quantum field theory; Choquet-Bruhat's focus was on
partial differential equations. She stayed in Princeton in 1951/52 and
also in 1955 when her father Gustave Choquet was likewise in the
Mathematics section at PIAS. Of course, Andr\'e Weil has spent a sizeable
part of his career there: from 1958 to 1976. Wether closer contacts
between him and Andr\'e Lichnerowicz have existed is unknown to
me. Also, I have no information about the relations between Princeton
University and the Sorbonne in Paris. According to D. Brill, it is
likely that Lichnerowicz has come through Princeton during the late
50s to early 60s.\\     

As to citations, Scherrer, Jordan, Thiry, and Jonsson could not have cited
Brans and Dicke. The only {\em symmetric} citation was Jordan
$\longleftrightarrow$ Thiry. We find a number of {\em one-way} citations,
e.g., Jordan $\rightarrow$ Jonsson; Scherrer $\rightarrow$ Ludwig \&
M\"uller; Jonsson $\rightarrow$ Thiry; Dicke $\rightarrow$ Jordan; Dicke \& Brans $\rightarrow$ Jordan; Brans $\rightarrow$ Jordan, Ludwig, Just. After the
papers of Brans and Dicke had appeared, those in Paris still working
on unified field theory took no notice of them. This refers to both, research on
Jordan-Thiry theory \cite{Surin1965}, and on Einstein-Schr\"odinger
theory \cite{PhoChaN1964}. Only M.-A. Tonnelat, in her books on the
experimental verifications of general relativity (1964), and on the
history of the relativity principle (1971), eventually cited both R. Dicke and
C. Brans. While in the second book they are mentionend only in
passing in connection with Mach's principle, (\cite{Tonnel1971},
p. 516), in the first one, Tonnelat did not give much weight to Brans-Dicke 
scalar-tensor theory: ``Recently, Brans and
Dicke [..] proposed a theoretical interpretation of the variation of
constants. [..] Actually, to us this hypothesis seems to be too
aleatory anyway to  here devote to it important theoretical
consequences'' (\cite{Tonnel1964}, p. 261).\footnote{R\'ecemment, Brans et 
Dicke [..] ont propos\'e une interpr\'etation th\'eorique de la variation des
  constantes. [..] Actuellement, cette hypoth\`ese nous semble
  n\'eanmoins trop al\'atoire pour que des d\'evelopments th\'eoriques
importants lui sont ici sp\'ecialement consacr\'es.} 

As to secondary literature, I have found only two citations of the second paper
by W. Scherrer \cite{Scherrer1949} - after the emergence of
Brans-Dicke theory. S. Deser and F. A. E. Pirani described Scherrer's
(and Dicke's) theory as ``in effect conventional versions of the
Hoyle-Narlikar theory.'' (\cite{DesPir1967}, p. 449).\footnote{Apart
  from the fact that this remark is a-historical, it also is a misunderstanding 
  of the intentions of both Scherrer and Dicke. For macroscopic
  phenomena, the Hoyle-Narlikar theory referred to is equivalent to Einstein's
  theory. Cf. (\cite{HoyNar1964}, pp. 198-200). For the difference
  between Hoyle-Narlikar and Brans-Dicke theory, see also \cite{Lord1976}, pp.
196-200).} E. S. Harrison in 1972 gave the same reference in a list of
papers concerning scalar-tensor theory without any comment \cite{Harris1972}.  

\section{Conclusions}

The previous discussion shows that the genesis of scalar-tensor theory
is not just a story about different people arriving independently at the same
theory: It is more intricate with some of the originators obtaining the same
formal theory and others a different one. The physical interpretations,
if available, also vary. During the process of theory-building, some of the 
researchers have come into contact with each other.\\ 
 
Now it is clear why Brans-Dicke theory
made such an impact. It was the slightest alteration of general
relativity of all scalar-tensor theories, and presented in a way that
everyone familiar with Einstein`s theory immediately could work with it
without having to learn new techniques. No strange five-dimensional spaces
around, no weird mass generation as a consequence as in Jordan's
original interpretation! No connection to unified field theory which
Thiry held high although it already carried the anathema of an
influential physicist of the time as W. Pauli. Certainly, some of the
physical ideas going into Brans-Dicke theory were either not new like
the conception of a variable gravitational constant. Or, they were on
shaky ground like the invocation of Mach's principle, ill defined for
a field theory and still living a meager life on the fringes of
physics. A new aspect with regard to Jordan was the discussion
concerning the weak and strong equivalence principle; it was most
important to Dicke because of his interest in devising and doing
experiments on gravitation. Up to this day, a possible violation of
the weak equivalence principle is the starting point for speculative
theory building. Jordan had to look for geophysical applications
(expansion of the Earth's crust) and to cosmology both of which were
not open to local experimentation. No doubt, the Lagrange function and
field equations, and the coupling with matter of Brans-Dicke theory
had been published years before by others. But Brans and Dicke opted
for the one Lagrangian, Fierz had shown to be irrefutable, and thus
they cleared up the clutter caused by the indeterminacy in the papers
of Jordan's coworkers Ludwig, M\"uller and Just.\\     

Coming to external factors for the immediate acceptance of
scalar-tensor theory \`a la Brans-Dicke: Princeton University with the
impressive figure of J. A. Wheeler in the community of relativists and
R. H. Dicke as a highly respected experimental and applied physicist, was
certainly one of the best places for authors to launch an alternative theory
of gravitation. Jordan's early papers and his books all were written
in German knowledge of which was decreasing rapidly after the
second world war. Also, some of his publications were hidden in special
journals, perhaps unavailable in most American libraries; in any case
they seemingly were not read abroad. At first, he had put into the foreground
his research on projective relativity, and he remained unclear with
regard to the coupling of the new scalar function to matter.\footnote{A further
  reason for reservation, in some circles in the United States, might have been 
Jordan`s nationalistic attitude during the Nazi regime. This is
mentioned by Carl Brans (\cite{Brans2008}, p. 5.)} As we have seen, 
interaction of the universities of Paris and the Institute for Advanced
Study in Princeton in the field of mathematics had been notable at the
time, but apparently there was very little exchange in terms of
physics. Also, the French scientists involved in unitary field theory
using Kaluza's five-dimensional theory, almost without exception
published in French.\footnote{Whether an influential figure like
  L. de Broglie with his negative stand toward quantum theory stood in
  the way of a stronger interaction between these two places is
  another question.} 

Willy Scherrer, a former president of  the Swiss Mathematical Society 
(1938-39), was the first one who had proposed a scalar-tensor theory
with formally the same Lagrangian Brans and Dicke were to use twenty
years later. Nevertheless, he was ignored by the community of relativists - his 
Swiss colleagues and P. Jordan included. In his applied-mathematics-attitude 
with regard  to scalar-tensor theory, he may be compared to
Y. Thiry, inclined more toward mathematical physics. Apparently,
Scherrer was not much respected by W. Pauli and M. Fierz and excluded
from their communication-lines.\\  

In comparison with the recent learned fights about the genesis of
general relativity, no high stakes were connected with scalar-tensor
theory. No battles about priorities 
ensued, no good anecdotes seem to have survived. Nevertheless, from
this case study we may learn how intimately connected speculative, empirical and
mathematical approaches to an alternative theory of gravitation have been.   
And, that communication between scientists in different countries was not as 
satisfactory as it might have been.
\section{Acknowledgment}
My thanks go to Dieter Brill for communicating some of his memories of
the time to me.

\end{document}